\documentclass[twocolumn,showpacs,preprintnumbers,amsmath,amssymb]{revtex4}
\usepackage{wasysym}
\usepackage{graphicx}
\usepackage{dcolumn}
\usepackage{bm}
\input epsf

\newcommand{\be}{\begin{equation}}
\newcommand{\ee}{\end{equation}}

\begin{document}

\preprint{}

\title{Energy transport in disordered classical spin chains}
\author{Vadim Oganesyan} \email{oganesyan@mail.csi.cuny.edu}
\affiliation{Department of Engineering Science and Physics, College of Staten Island, CUNY, Staten Island, NY 10314}
\author{Arijeet Pal}
\email{pal@princeton.edu} \affiliation{Department of Physics,
Princeton University, Princeton, NJ 08544}
\author{David A. Huse}
\email{huse@princeton.edu} \affiliation{Department of Physics,
Princeton University, Princeton, NJ 08544}
\date{\today}

\begin{abstract}
\noindent We present a numerical study of the diffusion of energy at high
temperature in strongly disordered chains of interacting classical
spins evolving deterministically. We find that quenched randomness
strongly suppresses transport, with the diffusion constant
becoming reduced by several orders of magnitude
upon the introduction
of 
moderate disorder.
We have also looked for but not found signs of a classical
many-body localization transition at any nonzero strength of the
spin-spin interactions.
\end{abstract}
\maketitle
\section{Introduction}
\label{sec:intro} Chemical, structural and other imperfections are
unavoidable in a typical crystalline solid. Often, reducing their
concentration reveals interesting intrinsic properties of the
ideal ``clean" material. It has long been recognized, however,
that sufficiently strong disorder can lead to a host of phenomena
essentially distinct from what is observed in clean materials.
Anderson localization \cite{pwa58} is perhaps the best known
example of such a phenomenon, whereby a quantum particle becomes
coherently trapped by the impurity potential and no longer
participates in transport. Essentially a wave phenomenon, Anderson
localization can also occur for classical linear waves, e.g. photons
or phonons.

It has often been assumed that true localization (strictly zero
diffusivity) of \emph{interacting} particles can only happen at
absolute zero temperature, even though Anderson's original paper on localization discusses
the possibility of localization persisting at nonzero temperature \cite{pwa58}.
Recently this question has been examined carefully by Basko and collaborators
\cite{baa}, who performed a stability analysis of an Anderson
insulator against weak interparticle interactions at low but
nonzero temperature. Their central conclusion is that an isolated
system of strongly disordered but weakly interacting quantum
particles should exhibit a transition into an insulating phase
with strictly zero diffusion at some low but nonzero excitation energy per particle (or temperature).
Motivated by this work, two of us \cite{OHED} considered quantum
lattice models that would be expected to exhibit such a dynamical
many-body localization transition as one varies the 
interactions or disorder strength, even at arbitrarily high temperature.  We attempted
to detect this transition by studying exact many-body spectral
statistics of small lattices. Initial results are encouraging
\cite{OHED}, although they fall short of making a strong case for
the existence of a quantum many-body localization transition, due
to strong finite-size effects.

Setting aside the question of the existence of such a transition
(i.e., assuming it does exist), one might wonder about its
nature, e. g., 
the universality class.
On the one hand, the theoretical analysis of Basko, {\it
et al.} \cite{baa} relies entirely on quantum many-body
perturbation theory.  Rather generally, however, one expects
macroscopic equilibrium and low-frequency dynamic properties of
interacting quantum systems at nonzero temperature to be
describable in terms of effective classical models.  This
expectation is certainly borne out in a variety of
symmetry-breaking phase transitions with a diverging correlation
length, such as, e. g., a
finite temperature N\'eel ordering of spin-1/2 moments. 
One can begin to understand the microscopic mechanism behind such
a many-body ``correspondence principle" as a consequence of an
effective coarse-graining, whereby the relevant degrees of freedom
are correlated spins moving together in patches that grow in size
as the phase transition is approached and therefore become
``heavy'' and progressively more classical. Further extension of
these ideas to general, non-critical, dynamical response is more
involved: roughly speaking, it requires that the typical many-body
level spacing in each patch be much smaller than the typical
matrix element of interactions with other patches. If this is true
(as it is in most models at finite temperature, though not
necessarily in the insulating phase analysed by Basko and
collaborators) one replaces microscopic quantum degrees of freedom
with macroscopic classical ones, which typically obey
``hydrodynamic'' equations of motion at low frequencies
\cite{subroto}.
Since it is expected that the many-body localization transition is
accompanied by a diverging correlation length (akin to the
Anderson transition) one might expect some sort of classical
description to emerge en-route from the localized phase to the
diffusive phase.  It was this thinking 
that initially motivated
us to consider the possibility of classical many-body
localization.

The process by which collective classical (hydro-) dynamics
emerges from a microscopic quantum description is subtle and may
or may not be relevant to the many-body localization discussed
above. A somewhat less subtle, but apparently largely unexplored
related question, is whether nonlinear, interacting and disordered classical many-body systems are capable of localization at nonzero temperature.
To be precise, a many-body classical dynamical system with a local
Hamiltonian
(including static randomness) 
should show hydrodynamic behavior, e.g. energy
diffusion, provided the local degrees of freedom are nonlinear and
interacting, and the disorder is not too strong.  In this regime,
the isolated system can function as its own heat bath and relax to
thermal equilibrium.  Diffusive energy transport must stop if the
interactions between the local degrees of freedom are turned off.
How is this limit approached?  Can there be a classical many-body
localization transition where the energy diffusivity vanishes
while the interactions remain nonzero?  These are the basic
questions we set out to investigate in this paper.

Our preliminary conclusion is that classical many-body systems
with quenched randomness and nonzero nonlinear interactions do generically
equilibrate, so there is no generic classical many-body localized
phase.  Our picture of why this is true is that generically a nonzero fraction 
nonlinearly interacting classical degrees of freedom are 
chaotic and thus generate a broad-band continuous
spectrum of noise.  This allows them to couple to and exchange
energy with any other nearby degrees of freedom, thus functioning
as a local heat bath.  Random classical many-body systems
generically have a nonzero density of such locally-chaotic
``clusters'', and thus the transport of energy between them is
over a finite distance and can not be strictly zero, resulting in
a nonzero (although perhaps exponentially small) thermal
conductivity.  Quantum systems, on the other hand, can not have a
finite cluster with a truly continuous density of states: the
spectrum of a finite cluster is always discrete.  Thus the
mechanism that we propose forbids a generic classical many-body
localized phase, yet it does not appear to apply to the quantum case.
The proposed existence of the many-body insulator in quantum problems is then a
remarkable manifestation of quantum physics in the macroscopic dynamics
of highly-excited matter.
In this paper we shall primarily focus on macroscopic low frequency behavior, postponing detailed analysis of local structure of noise and its relation to transport. Our conclussions are broadly consistent with findings of Dhar and Lebowitz\cite{dharlebowitz} although given the rather major differences in models, methods and, most importantly, the extend to which strongly localized regime is probed we refrain from making direct comparisons.

\begin{figure}[here]
\includegraphics[width=8.5cm]{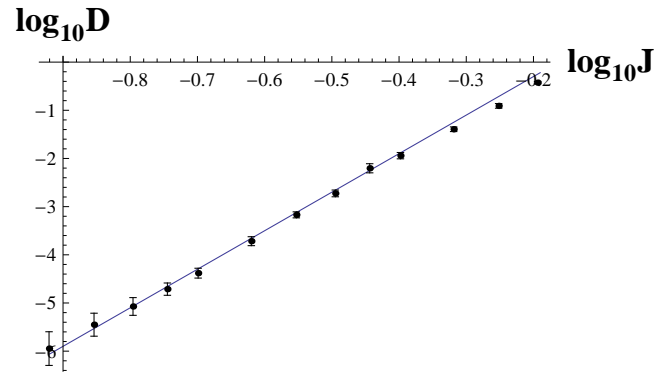}
\caption{Disorder-averaged energy diffusion constant $D$ as a function of
the spin-spin interaction $J$.
The line has slope 8 on this log-log plot. 
}
\label{fig:DofJ}
\end{figure}

We study energy transport in a simple model of
local many-body Hamiltonian dynamics
that has both strong static disorder and interactions: classical
Heisenberg spin chains with quenched random fields. For simplicity,
we consider the limit of infinite temperature, defined by
averaging over all initial conditions with equal weights.  Our
systems conserve the total energy and should exhibit energy
diffusion; they have no other conservation laws.  The energy diffusion coefficient,
$D$, can be deduced from the autocorrelations of the energy current (as
explained below) and is shown in Fig.~\ref{fig:DofJ} as a function
of the strength of the spin-spin interactions, $J$.  The
mean-square random field is $\Delta^2$, and as we vary $J$ we keep
$2J^2+\Delta^2=1$, as explained below.  The limit $J\to 0 $ is
where the interactions vanish, so there is (trivially) no energy
transport. 

As the interaction $J$ is decreased, the thermal diffusivity $D$ decreases very strongly; we have been
able to follow this decrease in $D$ for about 5 orders of
magnitude before the systems' dynamics become too slow for our
numerical studies.  
For most of this range, we can 
roughly fit $D(J)$ with a power-law, $D\sim J^\gamma$, with a
rather large exponent, 
$\gamma \cong 8$, as illustrated in Fig. 1.  
This large exponent suggests that the asymptotic behavior at small
interaction $J$ may be some sort of exponential, rather than power-law, behavior, 
consistent with the possibility that the
transport is actually essentially
nonperturbative in $J$. 
In principle, it is also possible to fit these data to a form with a 
nonzero critical $J_c$, so that $D(J<J_c)=0$  ---  such fits prove
inconclusive as they produce estimates of $J_c$ considerably
smaller than the values of $J$ where we can measure a nonzero $D$.
Since we are not aware of any solid theory for the behavior of $D(J)$,
these attempts at fitting the data are at best suggestive.
The large range of variation of the macroscopic diffusion constant $D$ across a rather modest range of $J$
is the most clearly remarkable and robust finding that we wish to present in this paper.

Our model and general methods employed will be presented and discussed in the next Section. Much of what we present is based on the analysis of energy current fluctuations in isolated rings. For various reasons we have found it beneficial to focus on these rather than fluctuations of the energy density or on current carrying states in open systems (we have spot-checked for quantitative agreement among these three methods). In Section III we present our results for macroscopic transport starting from short time behavior that is relatively easy to understand and working up to long time, DC behavior that is both difficult to compute and as of yet poorly understood. One particularly interesting observation we make here is that of a subdiffusive behavior at long times, apparently distinct from the much discussed mode-coupling behavior well representative of linear diffusion in the presence of disorder.
We discuss some afterthoughts and open problems in the Summary, with some important additional details relegated to the Appendices (such as quantitative explorations of finite-size effects, roundoff, many-body chaos and self-averaging).
\section{Model, trajectories and transport}
\label{sec:mtt} The classical motion of $N$ interacting particles
is usually defined by 
a system of coupled differential equations of motion. The
``particles" we study here are classical Heisenberg spins -- three-component
unit-length vectors, ${\bf S}_i$, 
placed at each site $i$ of a one-dimensional lattice.  With a
standard angular momentum Poisson bracket
and a Hamiltonian, $H$, the equations of motion are \be
\frac{\partial {\bf S}_j}{\partial t}={\bf H}_j\times{\bf S}_j,
\label{eq:eom} \ee where ${\bf H}_j=\partial H/\partial {\bf S}_j$
is the total instantaneous field acting on spin ${\bf S}_j$. The
Hamiltonians we consider are all of the form \be H=\sum_j({\bf
h}_j\cdot{\bf S}_j +J{\bf S}_j\cdot{\bf
S}_{j+1}),\label{eq:Hamiltonian} \ee with uniform pairwise
interaction $J$ between nearest-neighbor spins, and quenched
random magnetic fields, ${\bf h}_j$.  For almost all of the
results in this paper, we choose the random fields to be ${\bf
h}_j=h_j{\bf \hat n}_j$, where the $h_j$ are independent Gaussian
random numbers with mean zero and variance $\Delta^2$, while the
${\bf \hat n}_j$ are independent randomly-oriented unit vectors,
uniformly distributed in orientation.
Because of the random fields, total spin is not conserved and we
can focus on energy diffusion as the only measure of transport in
this system. For $J=0$ and $\Delta>0$ any initial distribution of
energy is localized, as the spins simply precess indefinitely
about their local random fields, so the diffusivity is $D=0$. In
the opposite limit, where $\Delta=0$ and $J>0$, there is diffusive
transport with $D\sim J$ (with nonlinear corrections due to the
coupling between energy and spin diffusion \cite{subroto,landau}).
We are interested in the behavior of $D$ as one moves between
these two limits,
especially as one approaches $J=0$ with $\Delta>0$. 

Given initial spin orientations, it is in principle
straightforward to integrate the equations of motion numerically,
thus producing an approximate many-body trajectory.  Correlation
functions can then be computed and averaged over a such
trajectories and over realizations 
of the quenched random fields. 
The transport coefficients can thereby be estimated via the
fluctuation-dissipation relations.
\subsection{The Model}
Before we embark on this program, however, we start by making a
change to the model's dynamics (\ref{eq:eom}), but not to its
Hamiltonian (\ref{eq:Hamiltonian}), in order to facilitate the
numerical investigation of the long-time regime of interest to us,
where the diffusion is very slow.  In order to get to long times
with as little computer time as possible, we want our basic time
step to be as long as is possible.  What we are interested in is
not necessarily the precise behavior of any specific model, but
the behavior of the energy transport in a convenient model of the 
type (\ref{eq:Hamiltonian}).  Since we are studying energy
transport, it is absolutely essential that the numerical procedure
we use does conserve total energy (to numerical precision) and
that the interactions and constraints remain local. Thus we modify
the model's dynamics to allow a large time step while still
strictly conserving total energy.

We change the equations of motion (\ref{eq:eom}) of our model so
that the even- and odd-numbered spins take turns precessing,
instead of precessing simultaneously.  We will usually have
periodic boundary conditions, so we thus restrict ourselves to
even length (thus bipartite) chains.  We use our basic numerical
time step as the unit of time (and the lattice spacing as the unit
of length). During one time step, first the odd-numbered spins are
held stationary, while the even-numbered spins precess about their
instantaneous local fields, \be
 {\bf H}_r(t)={\bf h}_r+J{\bf S}_{r-1}(t)+J{\bf S}_{r+1}(t)~,
 \label{eq:localfield}
 \ee
by the amount they should in one unit of time according to
(\ref{eq:eom}). Note that since the odd spins are stationary,
these local fields on the even sites are not changing while the
even spins precess, so that this precession can be simply and
exactly calculated, and the total energy is not changed by this
precession. Then the even spins are stopped and held stationary in
their new orientations while the odd spins ``take their turn''
precessing, to complete a full time step. Although this change in
the model's dynamics from a continuous-time evolution to a
discrete-time map is substantial, we do not expect it to affect
the qualitative long-time, low-frequency behavior of the model
that is our focus in this paper. In particular, we clearly observe
correct diffusive decay of local correlations for weak disorder
and essentially indefinite precession of spins at very strong disorder.

We have decided to use parameters
so that the mean-square angle of precession of a spin during one
time step is one radian (at infinite temperature), which seems
about as large as one can make the time step and still be roughly
approximating continuous spin precession.
This choice dictates that the parameters satisfy \be 2
J^2+\Delta^2=1~. \label{eq:JDelta} \ee   We will generally
describe a degree of interaction by quoting the $J$; the strength
$\Delta$ of the random field varies with $J$ as dictated by
(\ref{eq:JDelta}).
\subsection{Observables}
\label{subsec:observables} The basic observable of interest, the
instantaneous energy $e_i(t)$ at site $i$ is 
\be e_i(t)={\bf h}_i \cdot {\bf S}_{i}(t)+\frac{J}{2}({\bf
S}_{i-1}(t)\cdot {\bf S}_{i}(t)+{\bf S}_{i+1}(t)\cdot {\bf
S}_{i}(t))~. \ee  Note that 
with this definition, the interaction energy corresponding to a
given bond is split equally between the two adjacent sites.  When
updating the spin at site $i$, 
only the energies of the three adjacent sites, $e_i$ and $e_{i\pm1}$, change, due to the
change in the interaction energies involving spin $i$. 
This rather simple pattern of rearrangement of energy allows for
an unambiguous definition of the energy current at site $i$
during the time step from time $t$ to $t+1$. 
If site $i$ is even, so it precesses first, then the current is
\be j_i(t)=J[{\bf S}_i(t+1)-{\bf S}_i(t)]\cdot[{\bf
S}_{i+1}(t)-{\bf S}_{i-1}(t)]~, \ee  while for $i$ odd, \be
j_i(t)=J[{\bf S}_i(t+1)-{\bf S}_i(t)]\cdot[{\bf S}_{i+1}(t+1)-{\bf
S}_{i-1}(t+1)]~. \ee  

We are working at infinite temperature, or alternatively at
$\beta=(k_BT)^{-1}=0$.  The conventionally-defined thermal
conductivity vanishes 
for $\beta\rightarrow 0$ \cite{subroto}.
Instead, here we define the DC thermal
conductivity $\kappa$ so that the average energy current obeys \be
j=\kappa\nabla\beta \ee in linear response to a spatially- and
temporally-uniform small gradient in $\beta=1/(k_BT)$.  The Kubo relation
then relates this thermal conductivity at $\beta=0$ to the
correlation function of the energy current via \be
\kappa=\sum_{t}C(t)~, \ee where \be C(t)=\sum_i[\langle
j_0(0)j_i(t)\rangle]
\label{eq:kappa}
\ee is the 
autocorrelation
function of the total current, where the square brackets, $[\ldots]$, denote a
full average over instances of the quenched randomness
(``samples'') and the angular brackets, $\langle \ldots \rangle$, denote an average over initial
conditions in a given sample and time average within a given run. For our model (\ref{eq:Hamiltonian})
the average energy per site obeys \be \frac{d[\langle
e\rangle]}{d\beta}=\frac{J^2+\Delta^2}{3} \ee at $\beta=0$, and
the energy diffusivity $D$ is then obtained from the relation \be
\frac{\kappa}{D}=\frac{d[\langle e\rangle]}{d\beta}~. \ee

In a numerical study, if a quantity (such as $\kappa$) is
non-negative definite, then it is helpful to measure it if
possible as the square of a real measurable quantity.  We use this approach here, noting that \be
\kappa=\lim_{L,t\rightarrow\infty}\frac{1}{Lt}[\langle\{\sum_{\tau=1}^{t}\sum_{i=1}^Lj_i(\tau)\}^2\rangle ]~.
\ee  For a particular instance of the random fields in a chain of
even length $L$ with periodic boundary conditions and a particular
initial condition $I$ run for time $t$, we thus define the resulting
estimate of $\kappa$ as \be
\kappa_I(t)=\frac{1}{Lt}\{\sum_{\tau=1}^{t}\sum_{i=1}^Lj_i(\tau)\}^2~.
\label{eq:est} \ee  If these estimates are then averaged over
samples and over initial conditions for a given $L$ and $t$, this results
in the estimate $\kappa_L(t)=[\langle\kappa_I(t)\rangle]~.$  These
estimates $\kappa_L(t)$ must then converge to the correct DC
thermal conductivity $\kappa$ in the limits
$L,t\rightarrow\infty$.

\subsection{Finite-size and finite-time effects} 
\label{subsec:mttcomments} In a sample of length $L$, we expect
finite-size effects to become substantial on time scales \be t>t_L
= C_D L^2/D_{eff}, \label{eq:FSE} \ee where $D_{\rm eff}$ is the
effective diffusion constant at those time and length scales, 
and we find $C_D\cong 10$ (remarkably Eq. 15 remains valid more or less with the same value of $C_D$ across the entire range of parameters
-- see Appendix \ref{sec:FSE}).  With periodic boundary conditions
(which is the case in our simulations) this means that
$\kappa_L(t)$ saturates for $t>t_L$ to a value different from
(and usually above) its true DC value in the infinite $L$ limit, while with open boundary
conditions (no energy transport past the ends of the chain) the infinite-time limit of $\kappa_L(t)$
is instead identically zero for any finite $L$. 
We
simply avoid this purely hydrodynamic finite-size effect by using chains of large
enough length $L$, which is relatively easy, especially in the strongly-disordered
regime of interest, where $D_{eff}$ is quite small. 

For the smallest values of $J$ that we have studied, the system is
essentially a thermal insulator, and the $D_{eff}$ is so small 
that finite-size effects are just not visible at accessible times
even for small values of $L$, such as $L=10$. 
Instead, given the way we are estimating $\kappa$, a finite-time
effect, due to the sharp ``cutoffs'' in time at time zero and $t$
in (\ref{eq:est}), dominates the estimates $\kappa_L(t)\sim J^2/t$
in this small-$J$ regime. To explain this better we can rewrite the definition of $\kappa$ as
\be \kappa_L(t)=\frac{1}{Lt}\sum_{\tau=1}^{t}\sum_{\tau'=1}^{t}C(\tau-\tau')=\frac{2}{Lt}\sum_{t_{\rm av}=1}^{t/2} \kappa^*_L(t_{av}),
\ee
where we have assumed an even $t$ (there is an additional term otherwise) and $\kappa^*_L(\tau)\equiv\sum_{-\tau}^\tau C(\tau')$. Localization, i.e. zero DC conductivity, implies a rapidly vanishing $\kappa^*$ as well as $\kappa$ at long times. The latter however acquires a tail, $\kappa\sim 1/t$ whose amplitude is set by the short-time values of $\kappa^*$.

For the intermediate values of $J$ that are of
the most interest to us in this paper, there is also another,
stronger finite-time effect due to an apparently power-law ``long-time tail''
in the current autocorrelation function, $C(\tau)$, as we discuss in detail below. Importantly, at long times this intrinsic finite-time effect
dominates the extrinsic, cutoff-induced, $1/t$ effect discussed above, so
$\kappa_L(t)$ remains a useful quantity to study in this regime.
\section{results: macroscopic diffusion}
\label{sec:results}
\subsection{Current autocorrelations}
Since the total current is not dynamically stationary, its autocorrelation function, $C(t)$, should decay in time. 
In a strongly disordered dynamical system we expect the DC conductivity, which is the sum over all times of this autocorrelation function, to be very small due to strong cancelations between different time domains (i.e., $C(t)$ changes sign with varying $t$). The basic challenge of computing the DC thermal conductivity $\kappa$ boils down to computing (and understanding) this cancelation.

\begin{figure}[h!]
\includegraphics[width=8.5cm]{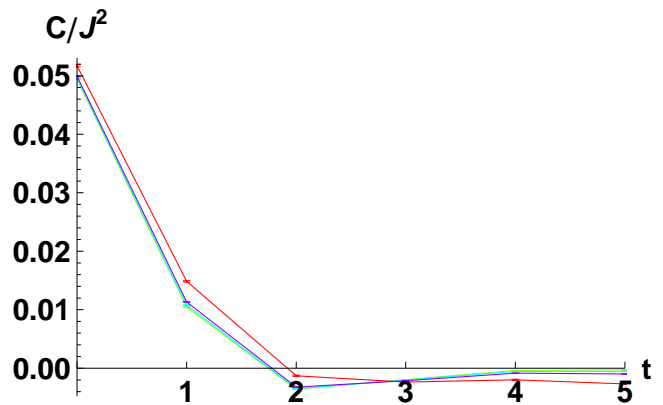}
\caption{(color online) Short time behavior of $C(t)$ for $J=0.32$ (red, noticeably different trace) and $J=0.08, 0.12, 0.16$ (these are almost identical data in this plot). Note rescaling of the vertical axis by $J^2$.}
\label{fig:shortC}
\end{figure}

\begin{figure}[h!]
\includegraphics[width=8.5cm]{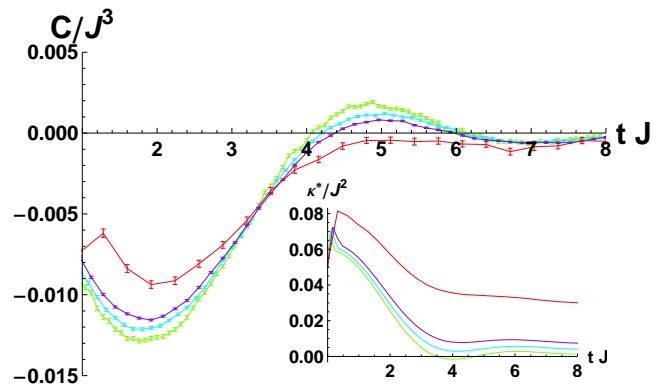}
\caption{(color online) Current autocorrelations on medium time scales $\sim 1/J$ for $J=0.32,0.16,0.12,0.08$, from top (red) to bottom (green) trace at $tJ=2$). Note the rescaling of both the vertical and time axes. The inset shows near cancellation between short and medium times.}
\label{fig:mediumK}
\end{figure}

The autocorrelation function $C(t)$ has three notable regimes as we
vary $J$ and $t$.  First, $C(t)$ is positive and of order $J^2$ at
times less than or of order one, as illustrated in Fig. \ref{fig:shortC}.
It quickly becomes negative at larger times.  For small $J$ it
is negative and of order $J^3$ in magnitude for times of order
$1/J$ (see Fig. 3).  For very small $J$, this negative portion of
$C(t)$ almost completely cancels the short-time positive portion,
resulting in an extremely small $\kappa^*$ (see inset in Fig. \ref{fig:mediumK}).
This cancelation is a hallmark of strong localization and can be observed, e.g. in an
Anderson insulator where it is nearly complete ($\kappa^*\to 0$ exponentially with time).
While the very short time behavior at small J is easily reproduced analytically by ignoring dynamical spin-spin correlations, the behavior out to times of order $1/J$ is representative of correlated motion of few spins (likely pairs). Although likely non-integrable, this motion is nevertheless mostly quasiperiodic  ---  we recorded indications of this in local spin-spin correlation functions (not shown here).

Finally,
there is apparently a power-law
long-time tail with a negative amplitude: $C(t)\sim -t^{-1-x}$, with
an exponent that we find is approximately $x \cong 0.25$ over an intermediate range of $0.2\lesssim J\lesssim 0.4$ (and more generally, perhaps).
\begin{figure}[h!]
\includegraphics[width=8.5cm]{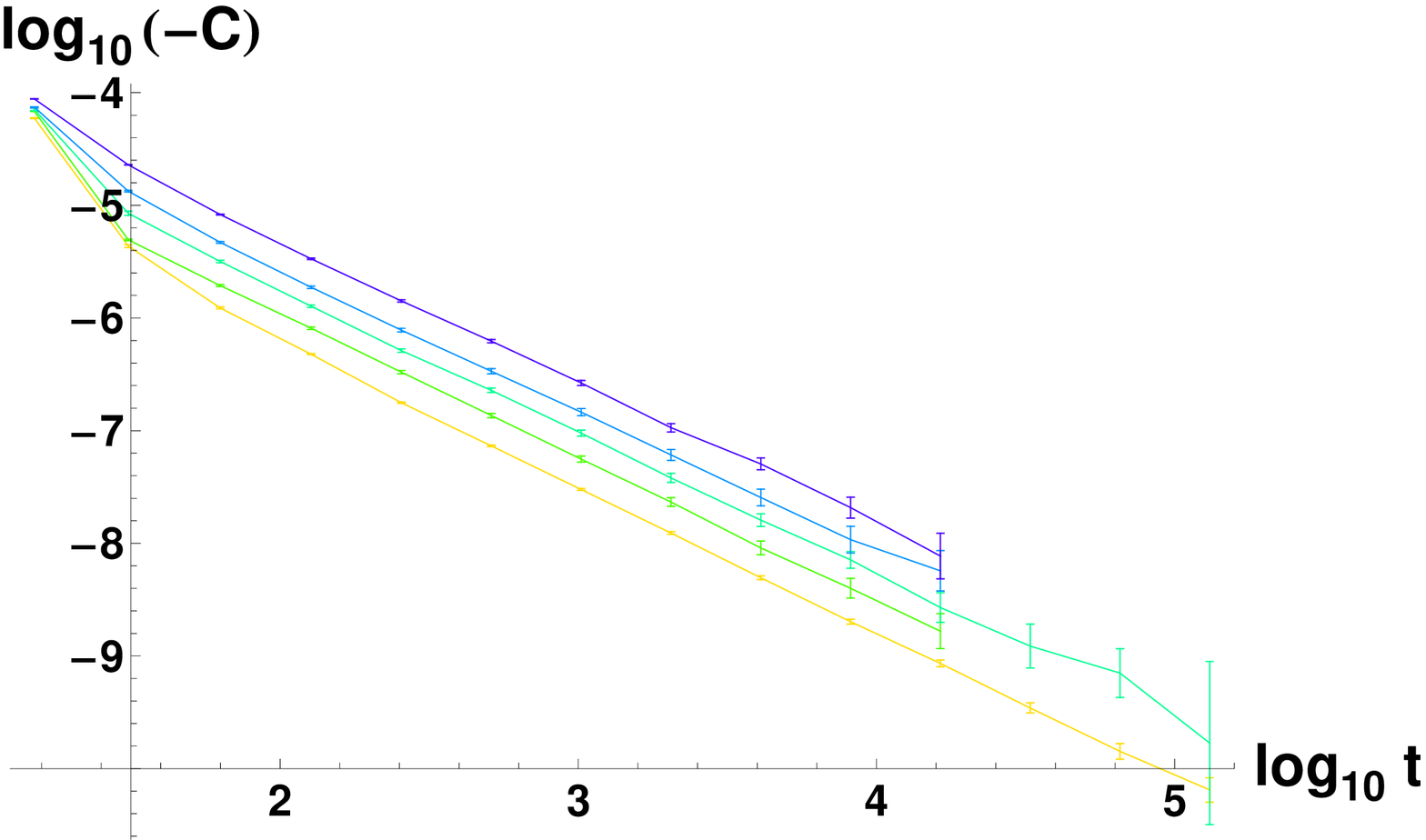}
\includegraphics[width=8.5cm]{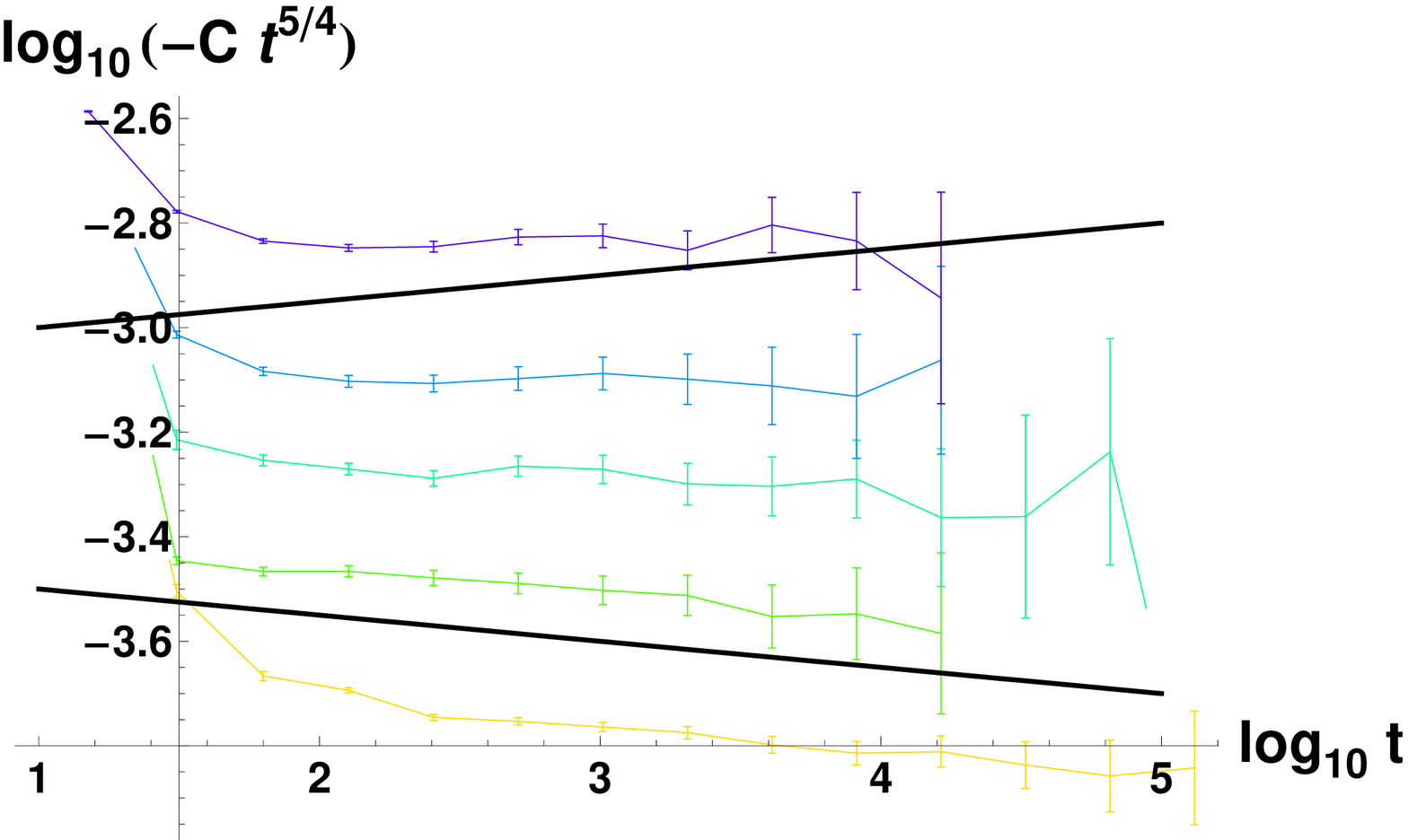}
\caption{(color online) Top panel: long-time tail in the current autocorrelation function for $J=0.20, 0.24, 0.28, 0.32, 0.40$ shown bottom to top in red, orange, yellow, green and blue, respectively. Bottom panel: to estimate the exponent we multiply the data by $t^{5/4}$ (and also display lines with slope $\pm 0.05$). Although these data do not exclude an exponent that varies with $J$, we interpret these results as supportive of a single exponent $x\approx 0.25$ at asymptotically long times but with a more pronounced short-time transient at smaller $J$.
}
\label{fig:longC}
\end{figure}
\begin{figure}[h!]
\includegraphics[width=8.5cm]{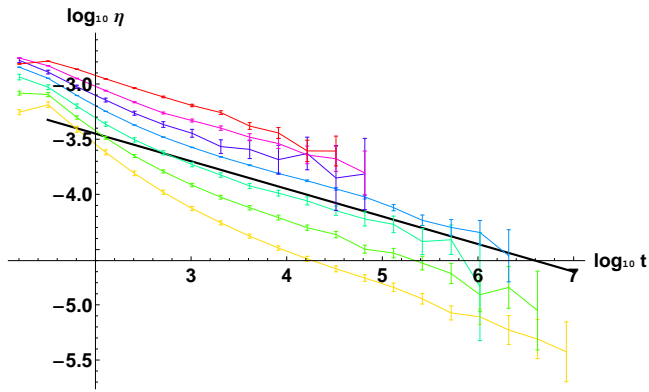}
\caption{(color online) Long time tails as seen from $\eta(t)$ for $J=0.20, 0.24, 0.28, 0.32, 0.36, 0.40, 0.48$ (bottom to top). Black line is a guide to the eye with slope $-1/4$.
Note that the short-time transients are stronger here, as compared to the auto-correlation data in Fig. 4.}
\label{fig:longDK}
\end{figure}
To observe this with the least amount of effort it is best to average $C(t)$ at long times over a neighborhood of $t$ (see Fig. \ref{fig:longC}) or to measure $\kappa_L(t)$ and compute its ``exponential derivative'', $\eta(t)\equiv\kappa_L(t)-\kappa_L(2t)$, at a sequence of points $t_n=2^n, n=1,2,3\ldots$ (see Fig. \ref{fig:longDK}).  The apparent value $x \cong 0.25$ of this exponent is something that
we do not understand yet theoretically.  However, we find that it does provide a good fit
to the data over a wide dynamic range, providing some support for our use of it to extrapolate to
infinite time and thus estimate the DC thermal conductivity, as discussed below.\subsection{DC conductivity: extrapolations and fits}
Our extrapolations of the DC conductivity will be based entirely on the long time behavior of
$\kappa_L(t)$ evaluated at a set of times $t_n=2^n$ with integer $n$ and for large enough $L$ to
eliminate finite-size effects (so we drop the subscript $L$).  We start by describing the procedure
used to arrive at the numerical estimates of the DC conductivity, then turn to the subject of uncertainties.

\begin{figure}
  \includegraphics[width=8.5cm]{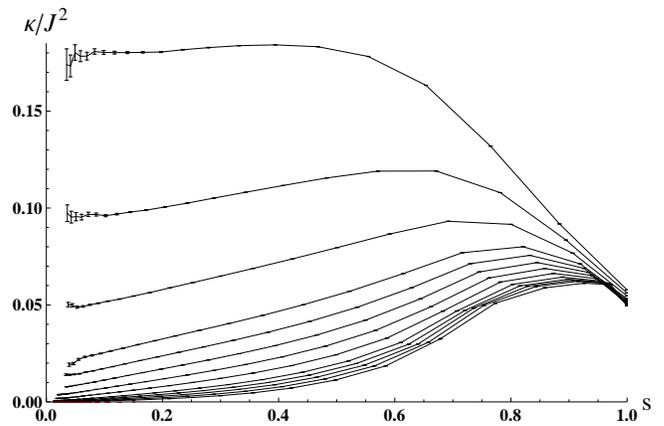}
  \caption{Variation of $\kappa(t)$ for 
$J=0.64, 0.56, 0.48, 0.40, 0.36, 0.32, 0.28, 0.24, 0.20, 0.18, 0.16,0.14,0.12$ plotted vs. $s=(1+Jt)^{-0.25}$.  Lines are merely guides to the eye, and statistical errors are too small to be seen on most of these points.
This figure is used for obtaining $J\geq0.36$ entries in Table \ref{table:vis}.}
\label{fig:vis1}
\end{figure}

A typical instantaneous value of the energy current is set by the strength of the exchange, $J$.  As a consequence $\kappa(t)\sim J^2$ for small $J$ at short and intermediate times ($t$ of order $1/J$ or less).  Given the time-dependence at intermediate and long times, as discussed above, we adopt the
variable $s=(1+Jt)^{-0.25}$ as a convenient ``scaling'' of time for displaying our results.
These rescalings ``collapse'' the observed values of $\kappa(t)$ for short to intermediate times across the entire range of $J$ studied, as shown  in Fig. \ref{fig:vis1}.

\begin{figure}
\includegraphics[width=8.8cm]{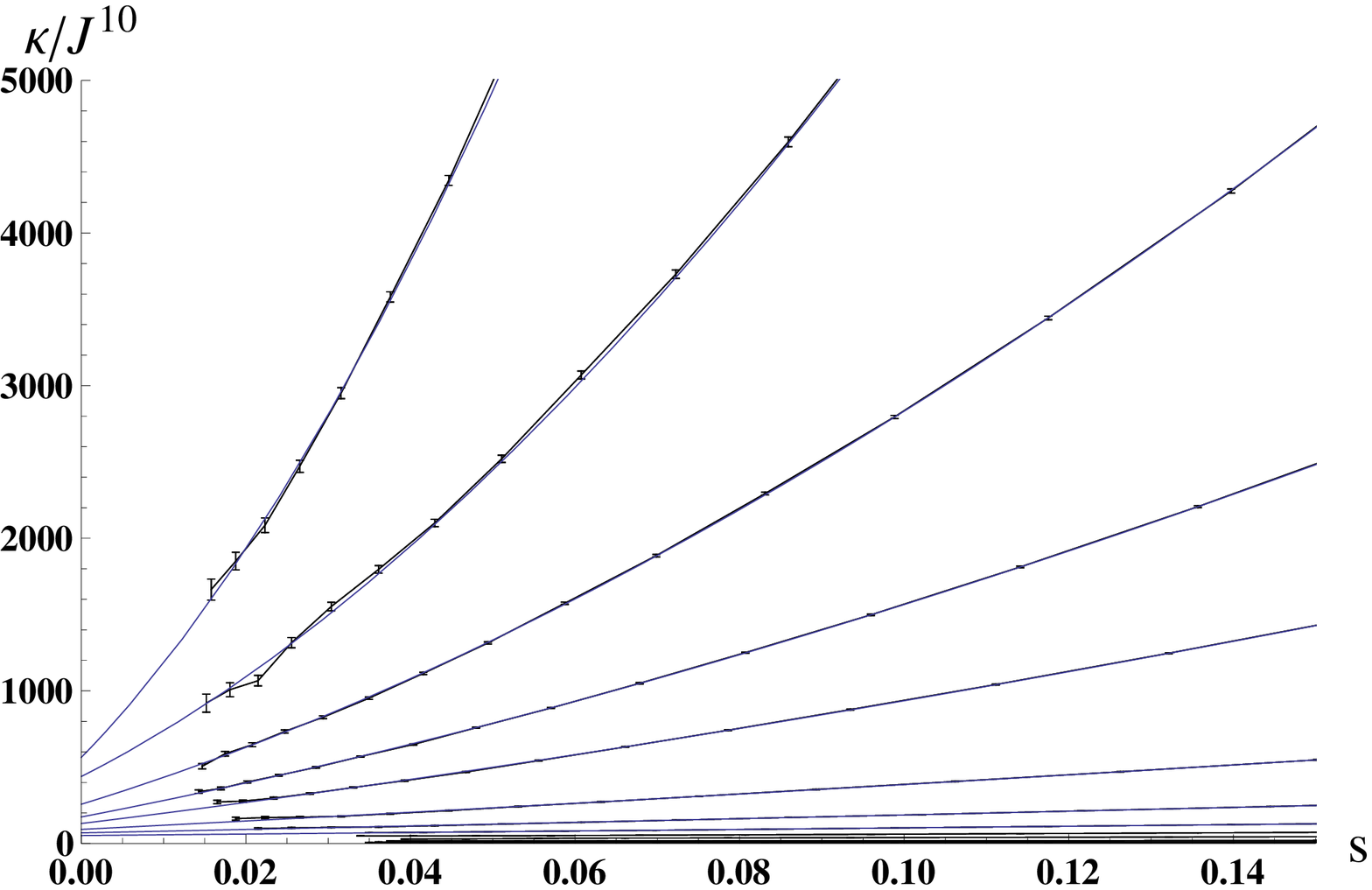}
\includegraphics[width=8.8cm]{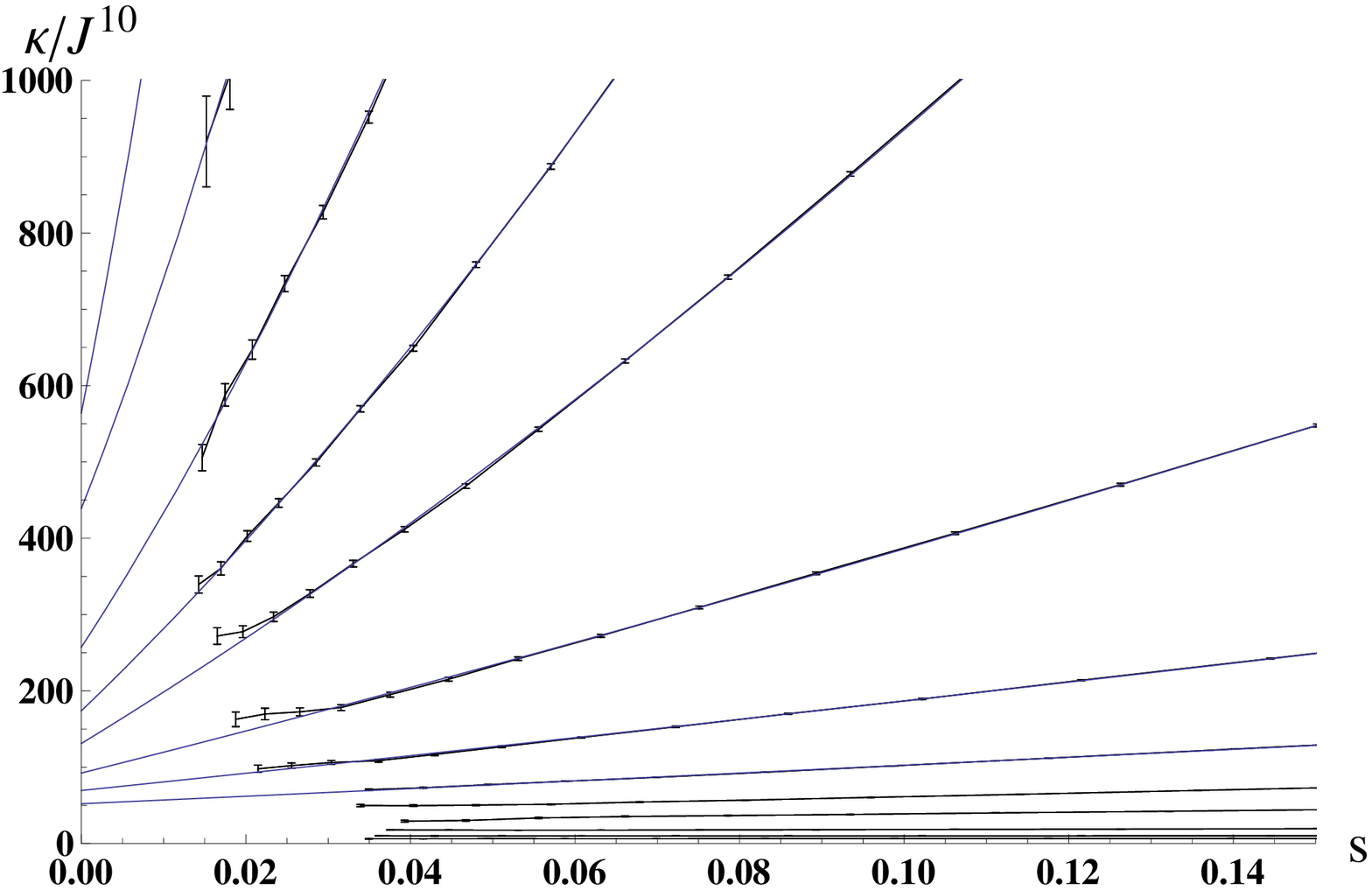}
\caption{(color online) Same data as in Fig. 6, but now scaled and displayed in a way
that allows one to see the extrapolations to $s=0$ ($t\rightarrow\infty$)
for small $J$.
Note the different rescaling schemes used in preceeding and current plots to focus on ``collapse'' of short time data (previous plot) vs. long time extrapolations (present plot). As before black lines are drawn through the data for guiding the eyes. Colored lines are results of polynomial fits, as explained in the text.  
This figure is used for obtaining entries in Table \ref{table:vis} for $J\leq0.32$.}
\label{fig:vis2}
\end{figure}

The extrapolated values of the DC conductivity decrease strongly as $J$ is reduced.  
Extrapolation of $\kappa(t)$ to $s=0$ and thus DC is fairly unambiguous for $J\ge 0.32$, 
as can be seen in Fig. \ref{fig:vis1}. 
To display the long-time results at smaller $J$, 
in Fig. \ref{fig:vis2} we  instead show $\kappa/J^{10}$.  Here one can see that as we go
to smaller $J$ the extrapolation to the DC limit ($s=0$) becomes more and more of
``a reach'' as $J$ is reduced.
The outcomes of these extrapolations and rough estimates of the uncertainties are summarized in Table \ref{table:vis}.
\begin{table}
\begin{center}
\begin{tabular}{
|c|c|c|c|c|c|}
	\hline
$J$ & $\kappa$& $\delta\ \kappa$
&$L$&$\log_2 T$&samples\\
\hline
0.64&$0.18 J^{2}$&$0.01J^{2}$
&5000&20&1000\\
0.56&$0.09 J^{2}$&$0.01J^{2}$
&2000&20&1000\\
0.48&$0.045 J^{2}$&$0.005 J^{2}$
&1000&20&4000\\
0.40&$0.020 J^{2}$&$0.003 J^{2}$
&1000&20&1000\\
0.36&$0.014 J^{2}$&$0.003 J^{2}$
&1000&21&2200\\
0.32&$50 J^{10}$&$8 J^{10}$
&1000&21&16000\\
0.28&$70 J^{10}$&$10 J^{10}$
&1000&24&912\\
0.24&$95 J^{10}$&$20 J^{10}$
&1000&25&558\\
0.20&$130 J^{10}$&$30 J^{10}$
&1000&26&1179\\
0.18&$175 J^{10}$&$50 J^{10}$
&500&27&2000\\
0.16&$250 J^{10}$&$100 J^{10}$
&500&27&1550\\
0.14&$400 J^{10}$&$200 J^{10}$
&500&27&520\\
0.12&$600 J^{10}$&$400 J^{10}$
&500&27&1116\\
	\hline
\end{tabular}
\caption{Extrapolated estimates of the DC conductivity $\kappa$, estimated uncertainties, length $L$ of samples,
and the number of time steps $T$ of the runs.
}
\label{table:vis}
\end{center}

\end{table}

There are several sources of uncertainty in the estimates of the DC thermal conductivity $\kappa$ 
reported in Table I. 
These can be separated into those originating with the measured values of $\kappa_L(t)$ and those due to the extrapolation to DC.

The statistical uncertainties in the measured values of $\kappa_L(t)$ were estimated (and shown in the figures) from sample-to-sample fluctuations which
we find follow gaussian statistics to a good approximation for these long (large $L$) samples.  We did look for a possible systematic source of error
originating with roundoff and its amplification by chaos (see Appendix B) 
and found it not to be relevant for the values of $J$ and $t$ studied.  

The uncertainties in our estimates of the DC $\kappa$ from the extrapolation procedure begin with the assumed value of the long-time powerlaw, $x \cong 0.25$.  Clearly,
using a different exponent will change the extrapolated DC values of $\kappa$ somewhat.  This 
uncertainty increases with decreasing $J$ as the ratio of the $\kappa_L(t)$ at the last time point to the extrapolated value increases.  At our smallest $J$ values, the
curvature in our $\kappa$ vs. $s$ plots due to the crossover to the earlier-time insulating-like $\sim 1/t\sim s^4$ dependence becomes more apparent and further complicates the extrapolation. Although we have experimented some with different schemes for extrapolating to DC, including different choices of exponent $x$, in the end the following procedure appeared to capture the overall scale of the diffusion constant, and with a generous estimate of the uncertainty: i) we start by removing early data with $s\gtrsim0.5$ to focus strictly on the long-time behavior; ii) this long time dependence is further truncated by removing 5 latest points and then fitted to a polynomial $\sum_0^4 a_n s^n$ to better capture the curvature apparent in the data -- these fits are shown in Fig. \ref{fig:vis2} and $a_0$ are the DC values reported in Table \ref{table:vis}; iii) the uncertainty is estimated as the greater of statistical error in the last point (which is negligible for most of our data) and the difference between $a_0$ and a simple linear extrapolation performed on latest five data points not included in (ii). 

Overall, we deem the values presented in Table \ref{table:vis} as ``safe'' since all extrapolated $\kappa$'s differ by at most a factor 2 from $\kappa$'s actually measured, in other words our extrapolations are reasonably conservative (with the exception of two smallest $J$'s where the extrapolations yields stronger reductions).
\section{Summary, further explorations and outlook}
In summary, we considered a rather generic model of classical Hamiltonian many-body dynamics with quenched disorder, and explored the systematic variation in the
thermal diffusivity between conducting and insulating states.  We found a rapid variation of the diffusion constant and presented quantitative estimates of the latter across more than $5$ orders of magnitude of change.
The origin of this behavior may be traced to spatial localization of classical few-body chaos -- we plan to present further results along these lines separately. Qualitatively, such a scenario is rather plausible at very low $J$, where most spins are spectrally decoupled due to disorder and essentially
just undergo independent Larmor precessions.  As long as $J$ is nonzero, however, there will always be a fraction of spins in
resonance with some of their immediate neighbors.  These clusters are then 
deterministically chaotic and
thus generate broad-band noise, which allows them to exchange energy with 
all other nearby spins. Importantly, in the entire parameter range studied
this heterogeneous regime eventually gives way at long time to a more homogeneous conducting state in the DC limit.
Thus, we suspect that internally generated but localized noise
always causes nonzero DC thermal transport 
even in the strongly disordered regime, as long as the spin-spin interaction $J$ is nonzero.

Additionally, we also discovered and characterized an apparent, novel finite-time (frequency) correction to diffusion,
with the diffusivity varying as $D(\omega)\approx D(0) + a |\omega|^x$ with $x\cong 0.25$.  Previous theoretical work on corrections to
diffusion due to quenched disorder\cite{ernst} have instead found a correction with exponent $x=1/2$, which is quite inconsistent with our numerical results.
This powerlaw behavior is apparently not due to the localization of chaos discussed above, as it persists well into the strongly
conducting regime (larger $J$) and also exists in models without a strong disorder limit at all
(e.g. with random fields of equal magnitude but random direction; data not shown).  So far we have not
found a theoretical understanding of these interesting corrections to simple thermal diffusion.
\begin{acknowledgments}
We thank J. Chalker, A. Dhar, S. Fishman, S. Girvin, J. Lebowitz, T. Spencer and especially S. Sachdev for useful discussions.
This work was supported by the NSF
through DMR-0603369 (V.O.) and MRSEC grant DMR-0819860 (D.A.H.). V.
O. was also supported by a Yale Postdoctoral Prize Fellowship.  A. P.
thanks the PCCM REU program at Princeton for support during the initial
stages of this work.
\end{acknowledgments}
\appendix
\section{finite size effects}
\label{sec:FSE}
\begin{figure}[here]
\includegraphics[width=8.5cm]{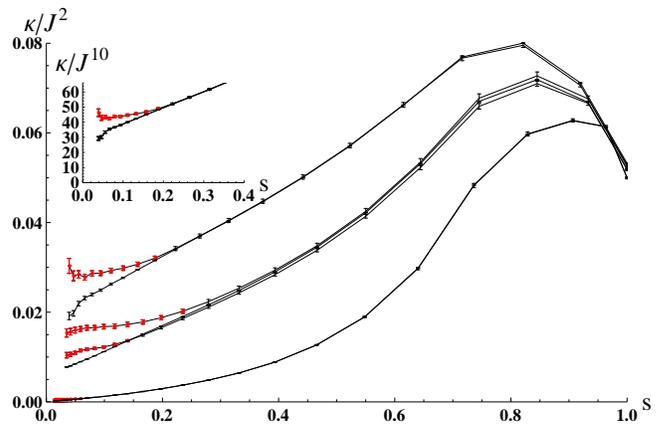}
\caption{(color online) Finite size effects for $J=0.16, 0.32,0.40$: 100 vs. 20 spins for J=0.40, 10 and 20 vs. 100 spins for J=0.32, 20 vs. 100 spins for J=0.16. Red color is used to indicate the data influenced by finite size effects according to Eq. ~\ref{eq:FSE} with $C_D=10$. Inset: $J=0.16$ data replotted.
}
\label{fig:FSE}
\end{figure}

We have checked that all of the extrapolations above are free from finite size effects (by comparing against simulations on smaller, and in some cases, larger samples). Nevertheless, it is interesting to consider the expected hydrodynamic size effects somewhat quantitatively, via Eq.~\ref{eq:FSE}. To illustrate this we display in Fig. ~\ref{fig:FSE} some results on shorter systems for $J=0.16, 0.32, 0.40$ that do show size effects: due to the periodic boundary conditions, the conductivity in smaller rings saturates in the DC limit at a value corresponding to the AC value at a ``frequency'' $2\pi/t^*$ corresponding to $\cong C_D D(2\pi/L)^2$, with $D\cong 3 \kappa$. Our results are qualitatively consistent this with $C_D\cong 10$ or slightly larger.  It perhaps remarkable that despite orders of magnitude of variation in the diffusion constant in going from $J=0.4$ to $J=0.16$ the crossover from bulk to finite system behavior is characterized by roughly the same constant $C_D\cong 10$. 
\section{Chaos amplification of round-off errors}
\label{subsubsec:Roff}  No numerical study of a nonlinear
classical dynamical system is complete without some understanding
of the interplay of discretization and round-off errors and chaos.
We are studying a Hamiltonian system that conserves total energy,
so the chaos is only within manifolds of constant total energy in
configuration space.  Thus although round-off errors introduce
tiny violations of energy conservation, these changes in the total
energy are not subsequently amplified by the system's chaos; we
have numerically checked that this is indeed the case. 
As a result of this precise energy conservation the energy
transport computation remains well-defined.  The simulation is far
less stable within an equal-energy manifold, where nearby
trajectories diverge exponentially due to chaos.  In particular,
this means that the component of any round-off error that is
parallel to the equal-energy manifold is exponentially amplified
by the chaos.  At large $J$ this happens rather quickly, while for
small $J$ the chaos is weaker and longer individual trajectories
can be retraced back to their respective initial conditions.
However, at small $J$ very long runs 
are necessary to extrapolate to the DC thermal conductivity: in
the end all of our extrapolations are done in the regime where all
individual many-body trajectories are strongly perturbed by
chaos-amplified round-off errors.

Ultimately, however, we are only concerned with the stability of
the
current autocorrelations that enter in the Kubo formula for $\kappa$. 
Although the precise trajectories may diverge due to chaos-amplified
round-off errors, this need not have a strong effect on $C(t)$.  To study this issue quantitatively we simulated roundoff
noise of different strength in our computations. Specifically, we
add extra random noise to the computation without altering the total energy by multiplying the angle each spin precesses in each time step by a factor of
$1+\eta_i(t)$,
where the
$\eta_i(t)$ are independent random numbers uniformly distributed between $P$ and $-P$ ($P$ = noise
strength).

In 400 rings of 500 spins coupled with $J=0.14$ we simulated the same initial
condition
with different $\eta_i(t)$, and with different values of simulated noise $P=10^0, 10^{-1}, 10^{-2}, 0 $ -- these results are
presented in Fig. \ref{fig:roff} below.
\begin{figure}
\includegraphics[width=8.5cm]{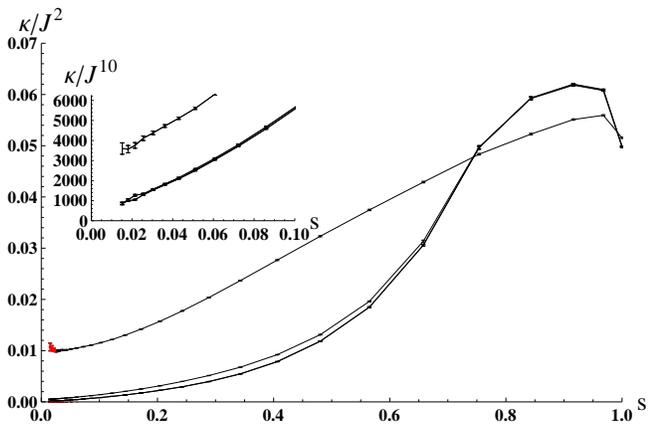}
\caption{
Roundoff effects at $J=0.14$: low frequency, long-time conductivity is larger for larger values of $P$ but is essentially indistinguishable between $P=0$ and $P=0.01$. Inset: data with $P=0.1, 0.01, 0$. 
}
\label{fig:roff}
\end{figure}
As expected, the long-time insulating behavior is weakened by the presence of noise.
Quantitatively, however, we observe little or no difference between results obtained
in the presence of simulated noise with $P=10^{-2}$ vs. ones obtained for intrinsic noise (which at double precision corresponds to $P_{intr}\lesssim 10^{-15}$). Clearly, this statement heavily depends on the duration of the simulation, value of $J$, etc.
Judging from Fig. \ref{fig:roff} roundoff errors are not a serious source of uncertainty in our results in main text. Interestingly, it is also possible for strong noise to suppress $\kappa$, as indeed happens at shorter times, which can be traced here to a sort of ``dephasing'' of sharp response of quasiperiodic localized states.


\end{document}